\documentclass[11pt]{article}
\newsavebox{\PSLASH}
\sbox{\PSLASH}{$p$\hspace{-1.8mm}/}

\begin{document}
\title{\large \bf On the stress-energy tensor of a null shell in Einstein-Cartan gravity}
\author{ S. Khakshournia$^{1}$
	\footnote{Email Address: skhakshour@aeoi.org.ir} and R.
	Mansouri$^{2,3}$ \footnote{Email Address: mansouri@ipm.ir}\\
	\\
	$^{1}$Nuclear Science and Technology Research Institute (NSTRI), Tehran, Iran\\
	$^{2}$Department of Physics, Sharif University of
	Technology, Tehran, Iran\\
	$^{3}$Institute for Studies in Physics and Mathematics (IPM), Tehran, Iran\\
}\maketitle
\[\]
\[\]

\begin{abstract}
 The stress-energy tensor of a matter shell whose history coincides with a null hypersurface in the Einstein-Cartan gravity is revisited.  It is 
demonstrated that with a proper choice for the torsion discontinuity  taken to be  orthogonal to the null hypersurface and consistent with the 
antisymmetric property of torsion tensor,  the modified expression for the asymmetric stress-energy tensor is automatically tangent to the hypersurface. 
The main differences with respect to a previous work are addressed.\\

\end{abstract}
\hspace{1.5cm}
\newpage
\section{Introduction}

The Einstein-Cartan (EC) theory is a natural generalization of general relativity that accounts for the presence of spacetime torsion which could be emanating from the spinning properties of the matter distribution. Hence, dynamics of the spacetime is determined by torsion being triggered by the intrinsic angular momentum of the matter, and the curvature being stemmed from the  mere presence of matter \cite{arcos2004torsion,blagojevic2013gauge,trautman2006einstein}.\\
The first generalization of the junction conditions for a null hypersurface within the EC gravity was made by Bressange \cite{bressange2000extension}(see also \cite{barrabes2003singular}). He extended the unified description of shells of any type, including the null case, provided by Barrabes-Israel \cite{barrabes1991thin} in the presence of a non-symmetric connection. In his approach the contorsion discontinuity, naturally appearing in the expression for the shell's stress-energy tensor, is taken to be orthogonal to the hypersurface, resulting in a modification to the surface stress-energy tensor. According to this modification, the expression for the stress-energy tensor of the shell is formally the same one as in general relativity but the tensor representing the jump of the transverse derivatives of the metric is replaced with a non-symmetric one splitting into a Riemann part and a Cartan part. Recently, using the Bressange's approach \cite{vignolo2018junction}, the junction conditions of two generic spacetimes through a non-lightlike hypersurface in the context of f(R) gravity with torsion has been derived via the definition of a suitable effective extrinsic curvature tensor which splits into a Riemann part and a Cartan one.\\  There is another approach to generalize the junction conditions for a singular hypersurface within the EC gravity applied to the Braneworld scenarios \cite{da2009braneworld}. Within this approach, it is the torsion discontinuity which is assumed to be orthogonal to the hypersurface. It turns out that  the expression for the shell's stress-energy tensor includes a term associated with the torsion sector of the connection. This term vanishes if we require that the surface stress-energy tensor be tangent to the hypersurface. As a result the Darmois-Israel junction conditions remain unchanged in the presence of torsion.\\
Motivated by the approach presented in \cite{da2009braneworld}, we assume an orthogonal torsion discontinuity across a null shell to find the surface stress-energy tensor in the presence of torsion. \\
\textit{Conventions.}  Natural geometrized units, $G=c=1$,  are used throughout the paper. The null hypersurface is denoted by
$\Sigma$. We use square brackets [F] to denote the jump of any quantity F across $\Sigma$. Latin indices range over the intrinsic coordinates 
of $\Sigma$ denoted by $\xi^{a}$, and Greek indices over the coordinates of the 4-manifolds. As we are going to work with distributional valued
tensors, there may be terms is a tensor quantity $F$ proportional to some $\delta$-function distribution. These terms are indicated by $\breve{F}$.\\

   \section{The Einstein-Cartan gravity}
In a Reimann-Cartan spacetime manifold, the torsion tensor is defined by the antisymmetric component of the affine connection as
   \begin{equation}\label{tortdef}
   	{T^{\sigma}}_{\mu\nu}={\Gamma^{\sigma}}_{\nu\mu}
   	-{\Gamma^{\sigma}}_{\mu\nu}.
   \end{equation}
Demanding that the metric tensor is covariantly constant, i.e. $ \nabla_{\sigma}g_{\mu\nu}=0 $, the following decomposition of the asymmetric connection can be done:
   \begin{equation}\label{tordecom}
   	{\Gamma^{\sigma}}_{\mu\nu}= \mathring{{\Gamma}^{\sigma}}_{\mu\nu}
   	+{K^{\sigma}}_{\mu\nu},
   \end{equation}
where $ \mathring{{\Gamma}^{\sigma}}_{\mu\nu} $ denotes the Christoffel symbols and $ {K^{\sigma}}_{\mu\nu} $ are the components of the contorsion or defect tensor of the connection given in terms of the components of the torsion by
   \begin{equation}\label{torcotdef}
   	K_{\sigma\mu\nu}=\frac{1}{2}(T_{\mu\sigma\nu}
   	+T_{\nu\sigma\mu} -T_{\sigma\mu\nu}),
   \end{equation}
   with $ K_{\sigma\mu\nu}=-K_{\mu\sigma\nu} $. The Einstein-Cartan field equations are
   \begin{equation}\label{torfieldeq1}
   	G_{\mu\nu}=8\pi T_{\mu\nu},
   \end{equation}
   \begin{equation}\label{torfieldeq2}
   	{T^{\mu}}_{\nu\sigma}+ \delta^{\mu}_{\nu} {T^{\rho}}_{\sigma\rho}- \delta^{\mu}_{\sigma} {T^{\rho}}_{\nu\rho}=8\pi {S^{\mu}}_{\nu\sigma},
   \end{equation}
   where $ {S^{\mu}}_{\nu\sigma} $ is the spin tensor representing the density of intrinsic angular momentum in the matter distribution related to the torsion tensor in a purely algebraic way. It must be noted that although Eq. (\ref{torfieldeq1}) is apparently identical to its general relativistic counterpart, here both $  G_{\mu\nu} $ and $ T_{\mu\nu} $ are generally asymmetric due to the presence of torsion.\\
   
    \section{The null shell formalism}
  
    Let $x^{\mu}$ be an admissible coordinate system in a coordinate neighborhood that includes the null hypersurface $\Sigma$ extending into both 
Riemann-Cartan spacetimes $\cal M^{\pm}$. The parametric equation of $\Sigma$ is written as $\Phi (x^{\mu})=0$, where $\Phi$ is a smooth 
function. The domains in which $\Phi$ is positive or negative are contained in $\cal M^{+}$ or $\cal M^{-}$, respectively. The metric across the two 
domains can be written as the distribution-valued tensor
 \begin{equation}\label{distmetric}
   g_{\mu\nu}=	g^{+}_{\mu\nu}\Theta(\Phi)+	g^{-}_{\mu\nu}\Theta(-\Phi),
\end{equation}
where $ \Theta(\Phi) $ is the step function and $ g^{-}_{\mu\nu} $ and $ g^{+}_{\mu\nu} $ are metrics in $ \cal M^{-} $ and $ \cal M^{+} $, continuously 
joined on $ \Sigma $. The corresponding tangent vectors on $ \Sigma $ are $ e_{(a)}=\partial/\partial\xi^{a} $, and a null normal vector is defined 
by $n_{\mu}=\alpha^{-1}\partial_{\mu}\Phi$, with $\alpha$ being a function on $\Sigma$, non-zero but otherwise arbitrary. 
The relevant jumps on $\Sigma$ expressed in the admissible coordinates $x^{\mu}$ must vanish:
 $[g_{\mu\nu}]=[n^{\mu}]=[N^{\mu}]=[\alpha]=0$. The completeness relations for the basis are written as \cite{barrabes1991thin}
 \begin{equation}\label{completerel}
            g^{\mu\nu}=g_{*}^{ab}e^{\mu}_{a}e^{\nu}_{b}- n^{\mu}N^{\nu}-n^{\nu}N^{\mu}.
      \end{equation}
In addition, we assume that the purely tangential part  of the torsion tensor be continuous across $ \Sigma $ \cite{bressange2000extension}:
       \begin{equation}\label{tortancomt}
       	e^{\mu}_{a}e^{\nu}_{b}e^{\sigma}_{c}[T_{\mu\nu\sigma}]=[T_{abc}]=0. 
       \end{equation}
The metric continuity condition implies that the tangential derivatives of the metric are continuous $[g_{\mu\nu,\sigma}]n^{\sigma}=[g_{\mu\nu,\sigma}]e^{\sigma}_{a}=0$. 
However, the transverse derivative of the metric, $g_{\mu\nu,\sigma}N^{\sigma}$, may be discontinuous, leading to
   \begin{equation}\label{metricdisc}
     [g_{\mu\nu,\sigma}] =-\gamma_{\mu\nu}n_{\sigma},	 
   \end{equation}
 where the tensor field $\gamma_{\mu\nu}$ is given by $ \gamma_{\mu\nu}= [g_{\mu\nu,\sigma}]N^{\sigma} $. Now the jump of the Christoffel symbols across $ \Sigma $ can be written as    
      \begin{equation}\label{Chrisdisc}
      	[{\stackrel{\circ}{\Gamma}}{}^\sigma_{\;\,\mu\nu}] =-
      	\frac{1}{2}(\gamma^\sigma_{\;\,\mu}
      	n_\nu+\gamma^\sigma_{\;\,\nu} n_\nu - \gamma_{\mu\nu}n^\sigma).
      \end{equation}
Taking into account the continuity of torsion tensor along $ \Sigma$ as expressed in (\ref{tortancomt}), any possible discontinuity must occur transverse 
to $ \Sigma $. Hence, we assume that the discontinuity of the torsion tensor across $ \Sigma $  can be written as
      \begin{equation}\label{tordisc}
       [{T^{\sigma}}_{\mu\nu}]=-\zeta_{\mu\nu}n^{\sigma},
      \end{equation}
for some  tensor $ \zeta_{\mu\nu} $ given by $ \zeta_{\mu\nu}=N_{\sigma}[{T^{\sigma}}_{\mu\nu}] $. Taking into account the antisymmetric property 
of ${ T^{\sigma}}_{\mu\nu} $ on the last two indices, the tensor $ \zeta_{\mu\nu} $ has to be antisymmetric. Then using Eq. (\ref{torcotdef}), the 
contorsion discontinuity takes the form \cite{da2009braneworld}
       \begin{eqnarray}\label{cotdis1}
           	[K^\sigma_{\;\,\mu\nu}] &=&
           -\frac{1}{2}(\zeta^{\sigma}_{\;\,\nu} n_\mu+\zeta^{\sigma}_{\;\,\mu} n_\nu
           	 -\zeta_{\mu\nu} n^{\sigma}).  
       \end{eqnarray}
Noting that $ \zeta_{\mu\nu} $ is an antisymmetric tensor, the expression (\ref{cotdis1}) is consistent with the antisymmetric property of the contorsion tensor on the first two indices.  
Eq.(\ref{cotdis1}) can be compared with the choice $[{K^{\sigma}}_{\mu\nu}]=\frac{1}{2}\beta_{\mu\nu}n^{\sigma} $, for some tensor $ \beta_{\mu\nu} $, made in the Bressange's approach \cite{bressange2000extension}. Based on Eqs. (\ref{Chrisdisc}) and (\ref{cotdis1}), we obtain from Eq. (\ref{tordecom}), the following expression for the jump of the asymmetric connection across $ \Sigma$:    
   \begin{eqnarray}\label{asymconjump}
   	[\Gamma^\sigma_{\;\,\mu\nu}] &=&
  -	\frac{1}{2}(\gamma^\sigma_{\;\,\mu}
        	n_\nu+\gamma^\sigma_{\;\,\nu} n_\mu - \gamma_{\mu\nu}n^\sigma)-\frac{1}{2}(\zeta^{\sigma}_{\;\,\nu} n_\mu+\zeta^{\sigma}_{\;\,\mu} n_\nu
        	-\zeta_{\mu\nu} n^{\sigma}).
   \end{eqnarray}
 The distribution-valued Riemann tensor is written as
   $ R^{\alpha}_{\;\,\mu\sigma\nu}={R^{+\alpha}}_{\mu\sigma\nu}\Theta(\Phi) +{R^{-\alpha}}_{\mu\sigma\nu}\Theta(-\Phi)
   +\breve{R}^{\alpha}_{\;\,\mu\sigma\nu}\delta(\Phi)$, where $ \alpha^{-1}\breve{R}^{\alpha}_{\;\,\mu\sigma\nu}=[{\Gamma^{\alpha}}_{\mu\nu}]n_{\sigma}-
      [\Gamma^{\alpha}_{\;\,\mu\sigma}]n_{\nu}$ \cite{barrabes2003singular}. Using Eq. (\ref{asymconjump}) we arrive at the following expression for the singular part of the Riemann tensor
      \begin{eqnarray}\label{Riemtensing}
         \alpha^{-1}\breve{R}^{\alpha}_{\;\,\mu\sigma\nu}&=&\left.\frac{1}{2}(\gamma^{\alpha}_{\;\,\sigma}	n_\mu n_{\nu} -\gamma_{\nu}^{\;\,\alpha}	n_\mu n_{\sigma} +
         	\gamma_{\mu\nu} n^\alpha n_\sigma-\gamma_{\mu\sigma} n^\alpha n_\nu)\right. \nonumber \\
         	&-&\frac{1}{2}(\zeta^{\alpha}_{\;\,\nu} n_\mu n_{\sigma}-\zeta^{\alpha}_{\;\,\sigma} n_\mu n_{\nu}-\zeta_{\mu\nu} n^{\alpha}n_{\sigma}+\zeta_{\mu\sigma} n^{\alpha}n_{\nu}).
           \end{eqnarray}
 The singular part $ \breve{R}_ {\mu\nu} $ of the Ricci tensor can be determined by contracting two indices in (\ref{Riemtensing}), yielding 
   \begin{eqnarray}\label{Ricctsing}
   \alpha^{-1}	\breve{R}_{\mu\nu}&=&\left.\frac{1}{2}(\gamma n_\mu n_\nu-\gamma_{\nu\sigma}
   n^\sigma	n_\mu-\gamma_{\mu\sigma} n^\sigma n_\nu)\right. -\frac{1}{2}(\zeta_{\mu\sigma} n_\nu n^{\sigma}-\zeta_{\nu\sigma} n_\mu n^{\sigma}),
    \end{eqnarray}
   and also the Ricci scalar
   \begin{eqnarray}\label{Riccssing}
   	\alpha^{-1}\breve{R} =
   \alpha^{-1}	g^{\mu\nu}\breve{R}_{\mu\nu} = -\gamma_{\sigma\nu} n^\sigma
   	n^\nu .
   \end{eqnarray}
   The singular part $ 	\breve{G}_{\mu\nu} $ of the Einstein tensor then takes the form 
   \begin{eqnarray}\label{Eintentsing}
   \alpha^{-1}	\breve{G}_{\mu\nu}&=&\frac{1}{2}(\gamma n_{\mu}n_{\nu}+g_{\mu\nu}\gamma_{\rho\sigma}n^{\rho}n^{\sigma}-\gamma_{\nu\sigma}
   	n^\sigma n_\mu-\gamma_{\mu\sigma}n^{\sigma}n_{\nu})\\ \nonumber
   &-& \frac{1}{2}(\zeta_{\mu\sigma} n_\nu n^{\sigma}-\zeta_{\nu\sigma} n_\mu n^{\sigma}).
   	\end{eqnarray}
   From the field equation (\ref{torfieldeq1})  the stress-energy tensor of the shell is written as\cite{poisson2002reformulation} 
   \begin{eqnarray}\label{surfaceengten}
   	8\pi S^{\mu\nu}&=&-\frac{1}{2}(\gamma n^{\mu}n^{\nu}+g^{\mu\nu}\gamma_{\rho\sigma}n^{\rho}n^{\sigma}-\gamma^{\nu}_{\;\,\sigma}n^\sigma n^\mu-\gamma^{\mu}_{\;\sigma}n^{\sigma}n^{\nu})\\ \nonumber
   	  &+& \frac{1}{2}(\zeta^{\mu}_{\;\,\sigma} n^\sigma n^{\nu}-\zeta^{\nu}_{\;\,\sigma} n^\mu n^{\sigma}). 	
   \end{eqnarray}
   In order to ensure that the asymmetric tensor $ S^{\mu\nu} $ is purely tangential to $ \Sigma $, the following conditions must hold:
   \begin{equation}\label{metricdisc}
   (i)S^{\mu\nu}n_{\mu}=0,	 \hspace{1cm}  (ii)S^{\mu\nu}n_{\nu}=0.
   \end{equation}
   It is easy to see that by virtue of the antisymmetric property of $  \zeta_{\mu\nu}$ two above conditions are automatically satisfied. The above 
expression for the stress-energy tensor of the null matter shell can be simplified after decomposing it into the basis of $ (N^{\mu},e^{\mu}_{a}) $. Using 
the completeness relations (\ref{completerel}), one can find the following decomposition \cite{poisson2002reformulation}: 
  \begin{eqnarray}\label{decompvect}
            (\gamma^{\nu}_{\;\,\sigma}- \zeta^{\nu}_{\;\,\sigma}) n^{\sigma}&=&\frac{1}{2}\gamma n^{\nu}-\frac{1}{2}\gamma_{ab} g_{*}^{\;\,ab}n^{\nu}+\zeta_{\lambda\sigma}N^{\lambda}n^{\sigma}n^{\nu}-\gamma_{\lambda\sigma}n^{\lambda}n^{\sigma}N^{\nu}\\ \nonumber
             &+& g_{*}^{\;\,ab} (\gamma_{\lambda\sigma}-\zeta_{\lambda\sigma})e^{\lambda}_{\;\,b}n^{\sigma}e^{\nu}_{\;\,a}.
    \end{eqnarray}
Inserting (\ref{decompvect}) into Eq. (\ref{surfaceengten}), and using once more the completeness relations together with a rearrangement of terms, we finally 
end up with the three-dimensional intrinsic form for the surface stress-energy tensor of the null shell in the presence of torsion:
 
   \begin{eqnarray}\label{stresstensor}
      16\pi	S^{ab}=-g_{*}^{cd}\gamma_{cd} n^{a}n^{b}-\gamma_{cd}n^{c}n^{d}g_{*}^{ab}
     +n^{a}g_{*}^{bc}(\gamma_{cd}-\zeta_{cd})n^{d} +n^{b}g_{*}^{ad}(\gamma_{cd}-\zeta_{cd})n^{c}.
   \end{eqnarray}
    From the expression (\ref{stresstensor}), the matter on the null shell can be characterized by 
  
   \begin{equation}\label{surfenergy}
  \sigma=-\frac{1}{ 16\pi}g_{*}^{cd}\gamma_{cd},
   \end{equation}
   as a surface energy density, and
    \begin{equation}\label{surfpress}
    p=-\frac{1}{ 16\pi}\gamma_{cd}n^{c}n^{d}\label{surfpres},
    \end{equation}
     as an isotropic surface pressure, and 
   \begin{equation}\label{surfcurnt1}
   j_{1}^{a}=\frac{1}{ 16\pi}g_{*}^{ad}(\gamma_{cd}-\zeta_{cd})n^{c},
   \end{equation}
    \begin{equation}\label{surfcurnt1}
    j_{2}^{b}=\frac{1}{ 16\pi}g_{*}^{bc}(\gamma_{cd}-\zeta_{cd})n^{d},
    \end{equation}
as the asymmetric surface energy currents. Note that due to the antisymmetric property of $ \zeta_{\mu\nu} $, one gets $ g_{*}^{cd}\zeta_{cd}=0 $ and $ \zeta_{cd}n^{c}n^{d}=0 $, indicating that the torsion contribution to the energy density $ \sigma $ in (\ref{surfenergy}) and isotropic surface pressure $ p $ in (\ref{surfpress}) vanishes, respectively.\\
Therefore, it is seen that in the presence of torsion, the surface stress-energy tensor of the null shell is modified as given in (\ref{stresstensor}). To compare this expression with that obtained in \cite{bressange2000extension}, we note that in the expression (67) in \cite{bressange2000extension} the torsion contribution to the stress-energy tensor encoded in the tensor $ \beta_{ab} $ associated with the jump of the contorsion across $ \Sigma $ appears with a positive sign, while here the same role played by the antisymmetric tensor $\zeta _{ab}$ associated with the jump of torsion across $ \Sigma $ turns out to have a negative sign.\\
It is seen from (\ref{stresstensor}) that there is a part  $\hat{\gamma}_{cd}-\hat{\zeta}_{cd} $ of $ \gamma_{cd}-\zeta_{cd} $ not contributing to  the matter content on the shell encoded in $S^{ab} $. This part satisfies the following 7 independent equations: 
 \begin{eqnarray}\label{stressindistmet3}
    	g_{*}^{cd}\gamma_{cd}=0,\\
    	(\gamma_{cd}-\zeta_{cd})n^{c}=0,\\ 
    	(\gamma_{cd}-\zeta_{cd})n^{d}=0 .
 \end{eqnarray} 
Since  $ \gamma_{cd}-\zeta_{cd} $ has 9 independent components, it follows that $\hat{\gamma}_{cd}-\hat{\zeta}_{cd} $ has two independent components  contributing to the Weyl tensor on the shell and can be interpreted as representing the two degrees of freedom of polarization of an impulsive gravitational wave traveling along the shell \cite{bressange2000extension}.\\
To explore the requirements on the spin tensor for generating the present discontinuity of torsion, 
we insert  (\ref{tordisc}) into the spin-torsion equation (\ref{torfieldeq2}), leading to
\begin{equation}\label{torfieldeq21}
	-\zeta_{\nu\sigma}n^{\mu}- \delta^{\mu}_{\nu} \zeta_{\sigma\rho}n^{\rho}+ \delta^{\mu}_{\sigma} \zeta_{\nu\rho}n^{\rho}=8\pi [{S^{\mu}}_{\nu\sigma}].
\end{equation}
Recalling the antisymmetric property of the spin tensor on the last two indices, the contraction of (\ref{torfieldeq21}) with the normal vector on each of the antisymmetric indices yields
 \begin{equation}\label{spinjump1}
	[{S^{\mu}}_{\nu\sigma}n^{\nu}]=0,\hspace{1cm} [{S^{\mu}}_{\nu\sigma}n^{\sigma}]=0,
\end{equation}
but the contraction with $ n_{\mu} $ leads to
\begin{equation}\label{spinjump2}
	8\pi[{S^{\mu}}_{\nu\sigma}n_{\mu}]=\zeta_{\nu\rho}n^{\rho}n_{\sigma}-\zeta_{\sigma\rho}n^{\rho}n_{\nu},
\end{equation}
indicating that while the contractions (\ref{spinjump1}) are continuous across the hypersurface, the antisymmetric tensor $ S^{\mu}_{\nu\sigma}n_{\mu} $ turns out to be discontinuous. A further contraction of (\ref{spinjump2}) with the normal vector yields
\begin{equation}\label{spinjump3}
 [{S^{\mu}}_{\nu\sigma}n_{\mu}n^{\nu}]=[{S^{\mu}}_{\nu\sigma}n_{\mu}n^{\sigma}]=0.
\end{equation}
The requirement (\ref{spinjump3}) on the spin tensor was also obtained in \cite{bressange2000extension} by imposing the condition that the surface stress-energy tensor must be tangential to $ \Sigma $, while in the present work it is unconditionally obtained. The intrinsic forms of the constraints (\ref{spinjump1}) are
\begin{equation}\label{spinjump4}
 [S_{abc}n^{b}]=[S_{abc}n^{c}]=0,
\end{equation}
while for (\ref{spinjump3}) it would be
\begin{equation}\label{spinjump5}
[S_{abc}n^{a}n^{b}]=[S_{abc}n^{a}n^{c}]=0.
\end{equation}
\section{Conclusion}
We have extended the Barrabes-Israel null shell formalism to the Einstein-Cartan theory of gravitation, assuming the torsion discontinuity to be orthogonal 
to the null hypersurface. The intrinsic stress-energy tensor in the presence of torsion is then modified and splits into a Riemann part and a Cartan part. To take into account the antisymmetric property of the torsion tensor on its last two indices, the tensor $ \zeta_{\mu\nu} $ associated with the torsion discontinuity is taken to be an antisymmetric tensor. As a result, the discontinuity of the contorsion tensor given by (\ref{cotdis1}) is in accord with its antisymmetric property and needs not to be added as a condition. In addition, the asymmetric stress-energy tensor of the null shell turns out to be unconditionally tangent to the  hypersurface $ \Sigma $.  Although the formalism was used for a null hypersurface, it may be applied to the nonlightlike cases accordingly.\\


\begin{thebibliography}{99}
	
\bibitem{arcos2004torsion}
  H. I. Arcos and J. G. Pereira, \textit{Torsion gravity: a reappraisal}, Int. J. 
  Mod. Phys. D, \textbf{13}, 2193-2240 (2004).
\bibitem{blagojevic2013gauge} 
  M. Blagojevic, F. W. Hehl, and T. Kibble, \textit{Gauge theories of gravitation: A
  reader with commentaries}, World Scientific (2013).
 \bibitem{trautman2006einstein}
  A Trautman, \textit{Einstein-Cartan theory}, Encyclopedia of mathematical physics, Editors J. P.
  	Francoise, G. Naber and S. Tsou, (2006).
 \bibitem{bressange2000extension}
  	G. F. Bressange, Class. Quantum Grav. \textbf{17}, 2509 (2000).
  	\bibitem{barrabes2003singular}
  	C. Barrabes and P. Hogan., \textit{Singular null hypersurfaces in general relativity: light-
  	like signals from violent astrophysical events}, World Scientific (2003).
\bibitem{barrabes1991thin}	
 C Barrabes and W Israel, Phys. Rev. D, \textbf{43}, 1129 (1991).
\bibitem{vignolo2018junction}
	S. Vignolo, R. Cianci, and S. Carloni, Class.  Quantum Grav. \textbf{35}, 095014 (2018).
\bibitem{da2009braneworld}
		J. M. Hoff da Silva and R. da Rocha, Class. Quantum Grav. \textbf{26},
		055007 (2009).
\bibitem{poisson2002reformulation}	
	E. Poisson, \textit{A reformulation of the barrabes-israel null-shell formalism}, \textbf{arXiv:gr-qc/0207101}.
\end{thebibliography}
\end{document}